# Effective conductivity in association with model structure and spatial inhomogeneity of polymer/carbon black composites

Z Garncarek†, R Piasecki‡, J Borecki‡, A Maj‡ and M Sudol‡

† Institute of Mathematics, University of Opole, Oleska 48, PL 45951 Opole, Poland
‡ Institute of Chemistry, University of Opole, Oleska 48, PL 45951 Opole, Poland

**Abstract.** The relationship between effective conductivity and cell structure of polyethylene/carbon black composites as well as between effective conductivity and spatial distribution of carbon black are discussed. Following Yoshida's model both structures can, in a way, be said to be intermediate between the well known Maxwell-Garnett (MG) and Bruggeman (BR) limiting structures. Using TEM photographs on composites with various carbon blacks we have observed that the larger is Garncarek's inhomogeneity measure $H$ of two-dimensional (2D) representative distribution of the carbon black, the smaller is the effective conductivity of the composite.

## 1. Introduction

Conductive polymer composites have found many applications in the electronics industry. In simple terms, a composite can be defined as a combination of several distinct materials, designed to achieve a set of properties not possessed by any of the components alone [1]. In the present paper we examine a composite which consists of two component materials—one forming a continuous phase (polymer matrix) and the other one forming discrete regions dispersed in the matrix (carbon black particles). The dependence of the electrical conductivity on the concentration of two species of carbon black is presented. To obtain structural information on the composites considered, the cellular effective medium approximation developed by Yoshida [2] is employed. To understand the electrical properties of the composites more clearly, knowledge of some features of the conductor–filler particle dispersion is essential [3]. Therefore, 2D spatial distributions of carbon blacks are characterized using Garncarek's novel method [4].

The paper is organized as follows. In section 2, the experimental procedure is described. In section 3, Yoshida's model used in this work is discussed briefly. In section 4, Garncarek's method is presented in outline and, on the basis of electron micrographs, is employed to evaluate the degree of inhomogeneity in the spatial distribution of carbon blacks. Finally, section 5 contains a discussion of the results and draws some conclusions.

## 2. The preparation of composite samples and electrical conduction measurements

As a basic polymer high-pressure polyethylene was used, produced by Blachownia, Kędzierzyn-Koźle. The Sakap 6 carbon filler (S6) was fabricated by Carbochem (Gliwice, Poland). The Printex XE 2 extra-conductive black (EC) was obtained from Degussa AG (Frankfurt, Germany). The composite components were mixed in a two-roller laboratory mill for 25 min at 423 K. Hereafter the composites will be referred to as P/S6 and P/EC for polyethylene/S6 and polyethylene/EC respectively. The composite samples with a different carbon content to 0.25 (0.35) by weight for EC (S6) were compressed under increasing pressures of 2 MPa/5 min, 10 MPa/3 min and 30 MPa/2 min to make films of 1 mm and of 2 mm in thickness. Current–voltage characteristics were evaluated at room temperature in a vacuum of about $1.3 \times 10^{-3}$ hPa. Taking into account the ohmic region the electrical conductivities of the composites were calculated. Low-conductivity samples were measured under 10 kV cm$^{-1}$ using a three-electrode sandwich-type system, while high-conductivity samples were measured under 10 V cm$^{-1}$ using a four-probe method to avoid a contact resistivity.

## 3. Application of Yoshida's model

The model introduced by Yoshida [2] (EMA-Y) to obtain the effective conductivity $\sigma^*$ of two-phase random composites, makes it possible to acquire some structural information in a particularly simple way. For that reason, the EMA-Y approach is used here. First let us start with a very brief outline of EMA-Y to get a better understanding of the meaning of the fitting parameters. Following Yoshida, one can generally observe three types of random composite geometry: (i) phase a (here carbon black with conductivity $\sigma_a$) surrounded by phase b (here polyethylene with conductivity $\sigma_P$); (ii) its inverse geometry; (iii) neither phase a nor phase b surrounded by a single phase alone.

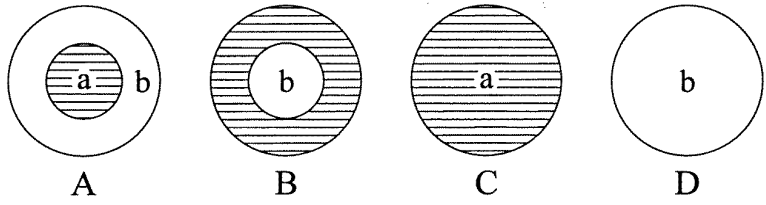

**Figure 1.** Four types of cell in Yoshida's approach [2]: complex cells A and B, and simple cells C and D.

The above mentioned structures are formed by mixing complex cells A and B (each consisting of core and shell) and the simple cells C and D (each consisting of a single phase), see figure 1. The former two indicate that a structural unit like a grain containing a single phase is surrounded by another phase alone, and the latter two show that such a unit is in contact with two phases arranged randomly around it. Three fitting parameters appear: $\phi$, $\alpha_1$, and $\alpha_2$. The parameter $\phi$ gives the volume fractions of the simple cells C and D. Parameters $\alpha_1$ and $\alpha_2$ denote carbon black volume fractions $x_v$ at which the complex cells B appear and the complex cells A disappear respectively. Using the effective medium concept to the assembly of cells, an equation for the effective conductivity $\sigma^*$ has been obtained (see equation (30) in [2]). In our notation this equation can be rewritten as

$$\frac{\sigma_a + 2\sigma_P + \gamma_A(\sigma_P - \sigma_a)}{(2\sigma^* + \sigma_P)(\sigma_a + 2\sigma_P) + 2\gamma_A(\sigma_P - \sigma^*)(\sigma_a - \sigma_P)} F_A$$
$$+ \frac{\sigma_P + 2\sigma_a + \gamma_B(\sigma_a - \sigma_P)}{(2\sigma^* + \sigma_a)(\sigma_P + 2\sigma_a) + 2\gamma_B(\sigma_a - \sigma^*)(\sigma_P - \sigma_a)} F_B$$
$$+ \frac{F_C}{2\sigma^* + \sigma_a} + \frac{F_D}{2\sigma^* + \sigma_P} = \frac{1}{3\sigma^*} \quad (1)$$

where $\gamma_K = \gamma_K(x_v, \alpha_1, \alpha_2)$ denotes the ratio of the cell and the core radii of the complex cells of type $K$ ($K = A, B$), and $F_i$ are the proper volume fractions of cells of type $i$ ($i = A, B, C, D$) depending on $x_v, \phi, \alpha_1, \alpha_2$ or $x_v, \phi$ for cells A, B or C, D respectively. For more details the reader is referred to Yoshida's original paper [2]. With reference to carbon black fillers the model considered has previously been employed to examine the composition dependence of the electrical and thermal conductivities as well as to the thermopower of copper phthalocyanine/Sakap 6 mixtures [5, 6]. Since, generally, the equation for $\sigma^*$ cannot be tackled analytically we have solved it on a computer. To convert the carbon black weight fractions $x_w$ to their volume fractions $x_v$ the following relationship is used:

$$x_v = 1 - (1 - x_w)g(x_w)/g_P \quad (2)$$

where $g_P = 0.916$ g cm$^{-3}$ is the polyethylene density and

$$g(x_w) = 0.324 x_w + 0.916 \quad \text{for (P/S6)} \quad (3a)$$
$$g(x_w) = 0.380 x_w + 0.916 \quad \text{for (P/EC)} \quad (3b)$$

are the densities of the composites (see figure 2). As expected, the structure of both two-phase composites proved to be more complicated than can be described by the single limiting cases: $\phi = 0$ with $(\alpha_1, \alpha_2) = (0, 0)$ or $(1, 1)$ for the MG theory [7] applicable to two-phase composites with a separated grain structure and $\phi = 1$ regardless of $\alpha_1, \alpha_2$ for the BR theory [8] applicable to composites with an aggregate structure. In our case the best fits obtained are

$$\phi = 0.6200, \alpha_1 = 0.0326, \alpha_2 = 0.5310$$
$$\sigma_a \equiv \sigma_{S6} = 2.14 \text{ S cm}^1 \text{ for P/S6} \quad (4a)$$
$$\phi = 0.1255, \alpha_1 = 0.0138, \alpha_2 = 0.3243$$
$$\sigma_a \equiv \sigma_{EC} = 12.08 \text{ S cm}^{-1} \text{ for P/EC} \quad (4b)$$

(see figure 3). Both $\phi$ values suggest some intermediate type of composite geometry, i.e. between MG and BR structures. However, for P/EC with $\phi$ much smaller than 0.5, the composite geometry is similar to the MG case, while for P/S6 the value of $\phi$ implies a mixed composite geometry. Therefore, for P/EC, relatively higher fractions of complex cells (characteristic for MG) over simple cells (typical for BR) in comparison with P/S6 are observed in figures 4(a) and 4(b). On the other hand the fitting parameter sets of $(\alpha_1, \alpha_2)$ indicate the following.

(1) If $0 < x_v < 0.0326$ ($0 < x_v < 0.0138$) for P/S6 (P/EC), then there is no B cell and the rest of the cells appear in the proper volume fractions $F_A$, $F_C$, and $F_D$ (for more details the reader is referred to Yoshida's original paper [2]). Now type (i) of Yoshida's geometry dominates.

(2) If $0.0326 < x_v < 0.5310$ ($0.0138 < x_v < 0.3243$) for P/S6 (P/EC), then both complex cells A and B appear together with the simple cells C and D. This coexistence of the cells implies a tangled structure with two phases surrounding each other. This is type (iii) of Yoshida's geometry. Note that the tangled structure for P/S6 appears over a wider range of $x_v$ than for P/EC.

(3) If $0.5310 < x_v < 1$ ($0.3243 < x_v < 1$) for P/S6 (P/EC), then we have the inverse geometry to type (i), namely there is no A cell but the other cells do occur.

It is interesting to note that the structure observed [5, 6] in a copper phthalocyanine/carbon black mixture with fitting parameters $\phi = 0.583$, $\alpha_1 = 0.074$ and $\alpha_2 = 0.521$ is similar to the P/S6 structure; similar carbon black was used.

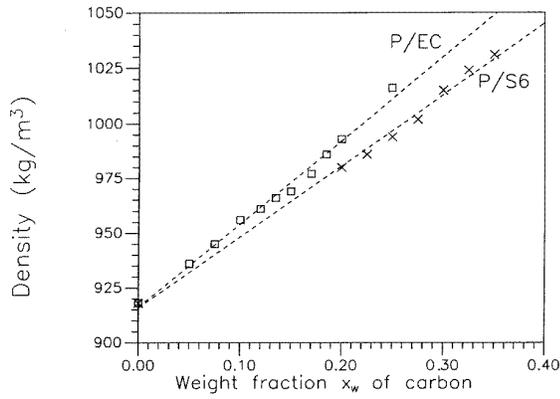

**Figure 2.** Density of composites as a function of carbon black weight fraction.

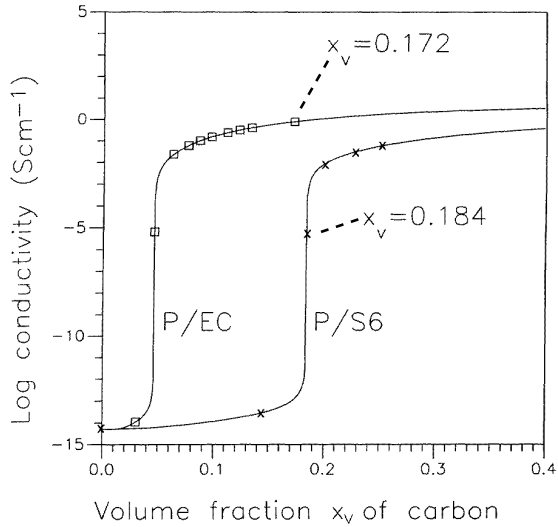

**Figure 3.** Effective conductivity $\sigma^*$ as a function of carbon black volume fraction for the composites. The curves are obtained from Yoshida's model using the parameters $\phi, \alpha_1$, and $\alpha_2$ given in the text by equations (4a) and (4b). The symbols □ and × denote experimental data.

## 4. Garncarek's quantitative criterion of inhomogeneity of the spatial distribution of carbon black in polyethylene

In recent papers [9, 10] an almost complete construction of the mathematical measure of the degree of inhomogeneity has been presented without proof. Proofs are given in the monograph [4], a small summary of which is given in the appendix. We shall only describe the basic concepts of the construction and final formulae here. Let us assume that we have some micrograph representing a real sample (see figure 5(a)). The transformation of data is illustrated in figure 5(b)–(d). First the micrograph is covered with an appropriate lattice (see figure 5(a)). Next, cells filled with a medium (no less than half of the cell area) in figure 5(a) are replaced by a cell containing a point as in figure 5(b). Each point in figure 5(b) is then replaced by the digit 1, while each empty cell is replaced by zero (figure 5(c)). Thus, an example of a population distribution chart (PDC) is obtained from figure 5(b). Figure 5(d) shows a second distribution chart of the same population, but now it is obtained on the greater scale $s/\kappa$, with larger unit cells, to cover figure 5(a). Here $s$ denotes an area of the micrograph expressed in terms of the smallest unit cell (now $s = 16$, see figure 5(c)) and $\kappa$ is equal to the number of unit cells of each chart ($\kappa = 16$ for figure 5(c) and $\kappa = 4$ for figure 5(d)). So we have two charts on increasing scales, the first one on 16/16 and the second on 16/4. In general, the family of PDCs consists of more charts since the first lattice covering the micrograph has a larger number of unit cells.

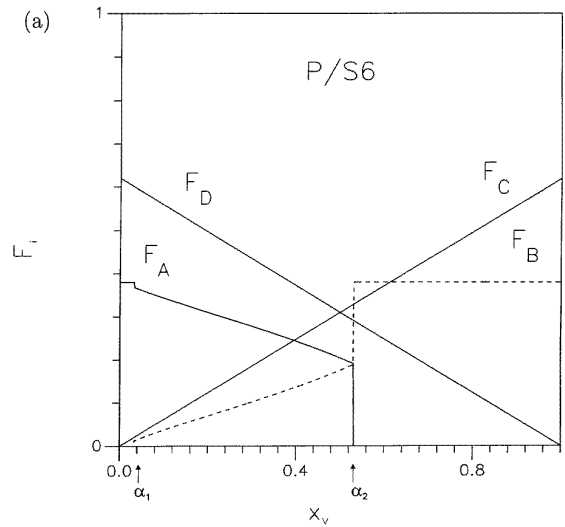

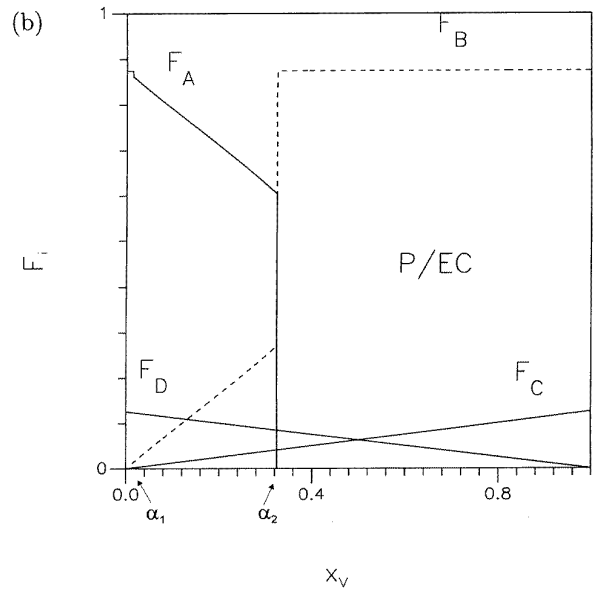

**Figure 4.** The proper volume fractions $F_A$, $F_B$, $F_C$, and $F_D$ of the cells of Yoshida's model [2] versus carbon black volume fraction. The structure parameters $\phi$, $\alpha_1$, and $\alpha_2$ are given in the text by equations (4a) and 4b). For clarity, broken lines are used for the $F$. (a) P/S6 composite. (b) P/EC composite.

Having finished the construction of the family of PDCs we come to the end of the data transformation process. In order to make a quantitative evaluation of

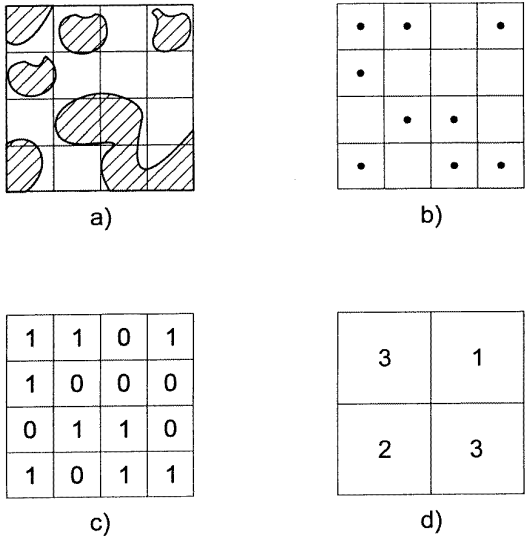

**Figure 5.** The transformation process of data. (*a*) An imaginary micrograph of nine objects covered with an appropriate lattice. (*b*) The points used on this lattice indicate cells filled with a medium. (*c*) The chart, on the scale 16/16 (see text), is obtained from the previous one using the following procedure: each of the dots is replaced by the digit 1, each of the empty cells by zero. (*d*) The chart of the same population but on the scale 16/4.

the agglomeration of carbon particles, a measure $h$ of the degree of inhomogeneity is introduced

$$h = \frac{\mu}{E(\mu)} \equiv -\frac{n}{\kappa - 1} + \frac{\kappa}{n(\kappa - 1)} \sum_{i=1}^{\kappa} n_i^2 \qquad (5)$$

where $\mu = \sum_{i=1}^{\kappa}(n_i - n/\kappa)^2$, $E(\mu)$ is the expected value of random variable $\mu$, $n_i$ is the number of objects in the $i$th cell of the PDC, $n = \sum_{i=1}^{\kappa} n$, and $\kappa$ denotes the number of unit cells; in the PDC (compare with equation (9) in [9]). The inhomogeneity index $h$ varies from $h = 0$, for a homogeneous distribution (in each unit cell of a chart $n_i = n/\kappa$), to $h_{max} = n$, for the case of maximal inhomogeneity (some $n_i = n$ and $n_j = 0$ for $j \neq i$). Using the measure $h$ one can always carry out a comparative analysis for a family of PDCs, but with the same or very similar numbers of objects [10]. It is interesting to note that some linear transformations of the configurational entropy and the measure $h$ are strongly correlated [11].

The three sigma test is often used in probability applications to evaluate the deviation of a random variable from its expected value. As has been already shown, an improved measure $H$ can be defined by

$$H = \frac{h - E(h)}{\sigma_h}$$
$$\equiv \left(\frac{n(\kappa - 1)}{2(n - 1)}\right)^{1/2} \left(\frac{1 - n - \kappa}{\kappa - 1} + \frac{\kappa}{n(\kappa - 1)} \sum_{i=1}^{\kappa} n_i^2\right) \qquad (6)$$

where $\sigma_h$ denotes the standard deviation of $h$ (we refer to original papers [4, 9] for more details). At present the possible values of the index $H$ belong to

$$\left\{-\left(\frac{n(n-1)}{2(\kappa - 1)}\right)^{1/2}, 0.5[2n(n-1)(\kappa - 1)]^{1/2}\right\}. \qquad (7)$$

The inhomogeneity index $H$ reaches negative values for a PDC with homogeneity features stronger than the random distribution. For a PDC with inhomogeneity features more distinct than a random distribution, the inhomogeneity measure $H$ attains positive values. In turn, for $H$ values near zero one can conclude that the distributions under consideration are random. Using the measure $H$ one can always perform a comparative analysis for a family of PDCs, even with significantly varying numbers of objects.

The above method is now applied to the polymer/carbon black composites. We utilized two different electron micrographs of 25 wt% P/S6 (final magnification factors of 12 000 and 47 000) and two different electron micrographs of 25 wt% P/EC at the same as above final magnifications. Working images were formed by extracting a square of $480 \times 480$ pixels from each digitized image. Note that our choice provides 23 different scales while the usually used square of $512 \times 512$ pixels gives only nine scales (the number of scales is equal to the number of divisions of the side length expressed in pixels). The digitized images were converted to black and white by choosing a greyness threshold for the pixels such that the area fraction determined from the picture matched the measured volume fraction of the specimen. According to relations (2), (3$a$) and (3$b$), $x_w = 0.25$ corresponds to $x_v = 0.172$ for P/EC and $x_v = 0.184$ for P/S6.

In this manner we obtained the two-dimensional projection (binary image) containing (indirectly) information about the three-dimensional distribution of carbon black in the thinnest layer of the composite with the proper volume fraction. So we can assume that we examined representative cross sections of the composites. These images are presented in figures 6($a$), ($b$), 8($a$) and 8($b$). For each considered image, a family of PDCs is formed. For the four families of PDCs obtained above, we calculate the values of the index $H$ in each considered scale. These curves are calculated as functions of unit cell side length $k$ expressed in pixels, $k = 480/\sqrt{\kappa}$, and are plotted in figures 7 and 9.

## 5. Discussion of results and conclusions

Regarding optical properties [12] one can state that there is no universal solution to modelling the electrical conductivity of binary composites, since the effective properties depend on the spatial distribution of the conductive phase in the host medium. Similarly, the effects of agglomeration of carbon particles in the semiconducting material on the dielectric strength of polyethylene insulation have been studied [13]. On the basis of relating the average auto-correlation function to darkness profile along a line on the micrograph, the authors found that the agglomeration of acetylene black particles in the semiconducting material ranges from 100 to 300 nm. They suggested that proper control of the agglomeration of the carbon particles by mixing some additives could improve the minimum breakdown strength of power cables for EHV transmission. This is an example of 2D data analysis to obtain a possible explanation of the 3D problem.

In our case the cross sections mentioned above are assumed to be representative for 3D composites with a given carbon black fraction. We consider the reproducible samples obtained with the same processing conditions. Hence, similar spatial distributions of carbon black (characteristic of a given type of carbon black) should appear. Therefore the dispersion of values of $H$ should be relatively very small on each scale under consideration. Let us return to figures 7 and 9. In both figures the examined curves show a very large degree of spatial inhomogeneity. As expected, with increasing magnification (in our case from $\times 12\,000$ to $\times 47\,000$) $H$ also increases because there are carbon black aggregates which are relatively large with regard to the area of image. It is interesting to note that for comparable fractions of various carbon blacks (in our case for the fractions $x_v = 0.172$ (EC) and $x_v = 0.184$ (S6)) the following effect can be observed: *the larger the inhomogeneity measure $H$ of a 2D representative distribution of carbon black, the smaller is the effective conductivity of the composite.* In our case the difference in the carbon black conductivities is about 10 S cm$^{-1}$, giving about 20% of the difference observed in figure 3. It seems most likely that the inequality $\sigma(\text{P/EC}) > \sigma(\text{P/S6})$, particularly distinct for the range $0.05 < x_v < 0.20$ (see figure 3), can be linked mainly with the inequality $H(\text{P/EC}) < H(\text{P/S6})$ caused by different spatial distributions of carbon black independently of the magnification factor.

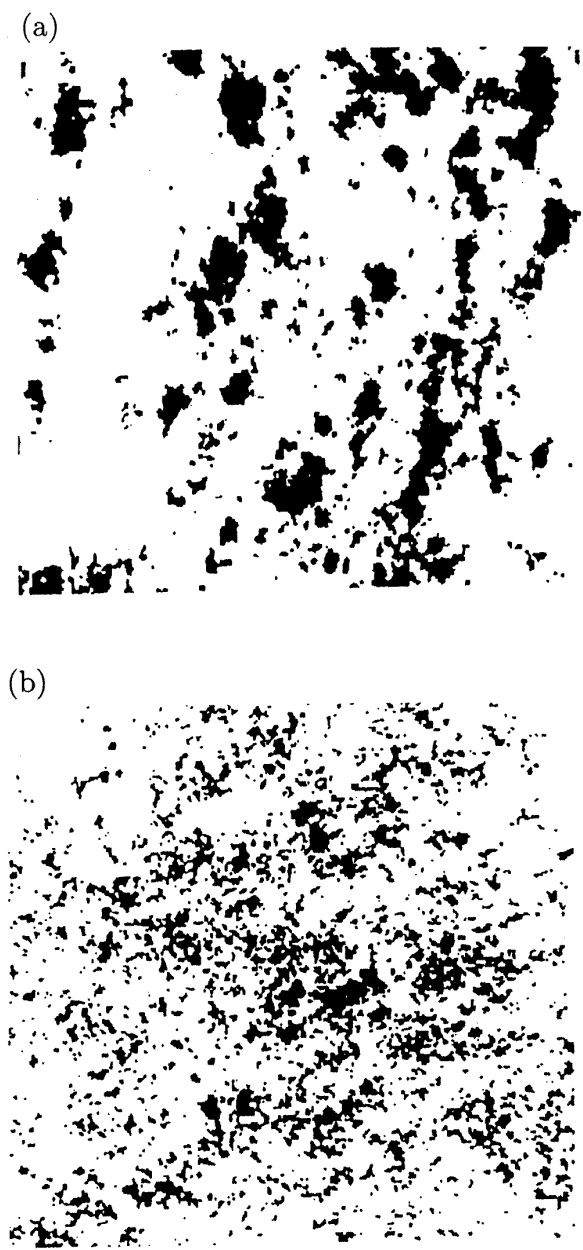

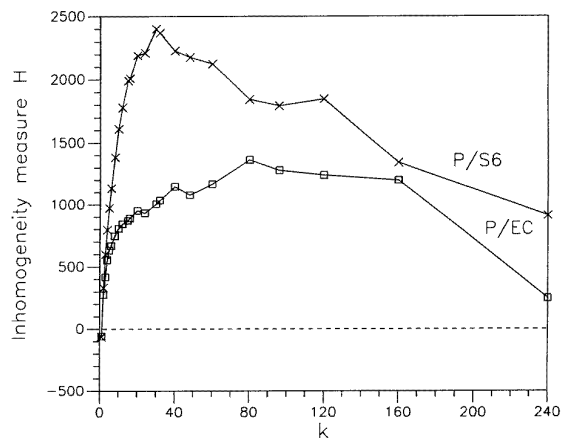

**Figure 7.** A plot of the measures $H(\text{P/S6})$ and $H(\text{P/EC})$ for a magnification of $\times 12\,000$ as functions of different scales $s/\kappa$, where $k = 480/\sqrt{\kappa}$. The broken line represents the level of ideal randomness, $H = 0$.

**Figure 6.** Representative binary images, $480 \times 480$ pixels, of composites at a magnification of $\times 12\,000$. (*a*) P/S6 composite. (*b*) P/EC composite.

Considering together the results of Yoshida's model and Garncarek's method, we can vouch for consistency between 2D representative distributions of carbon blacks and the 3D effective conductivity of the composites. For instance, the experimental effective conductivity of the P/EC composite shows its critical volume fraction is much lower than for the P/S6, see figure 3. The theoretical curves $\sigma^*$ obtained from Yoshida's model fit the experimental data with essentially different sets of parameters. One of them, the parameter $R$, allows us to distinguish geometrical structures of composites. It takes very different values for P/EC and P/S6. This is in agreement with very different 2D representative distributions of various carbon blacks in figure 6(*b*) and 6(*a*) respectively. The easily seen more uniform distribution, but still with strongly ramified clusters, for the P/EC is quantitatively characterized on different scales by the inhomogeneity index $H$ with much lower values than for P/S6 as expected. The most likely interpretation is that the higher homogeneity of the 2D representative distribution for P/EC favours the enhanced 3D interconnectivity of the carbon particles. Obviously, this factor is a positive influence on the electrical conductivity of P/EC in agreement with earlier findings about its critical volume fraction.

Recently computer simulations of the spatial disorder dependence of the conductance of a simplified Miller–Abrahams random resistor model showed that the resistance of 3D samples decreases with the spatial disorder [14]. However, the authors introduced a totally different measure of spatial disorder which involves only one-dimensional displacements of all the sites. So this measure represents some sort of projection of *n*-dimensional data $(x_i, y_i, z_i)$ onto one dimension $(x_i)$ and works at one scale only. In our case, Garncarek's measure $H$ was calculated for a family of PDCs, i.e. for many scales. Moreover, $H$ can be calculated for an *n*-dimensional distribution if we know *n*-dimensional data. In this paper we have restricted ourselves to calculation of the index $H$ for two-dimensional representative distributions of carbon black because at this stage the only available data are provided by transmission electron microscopy.

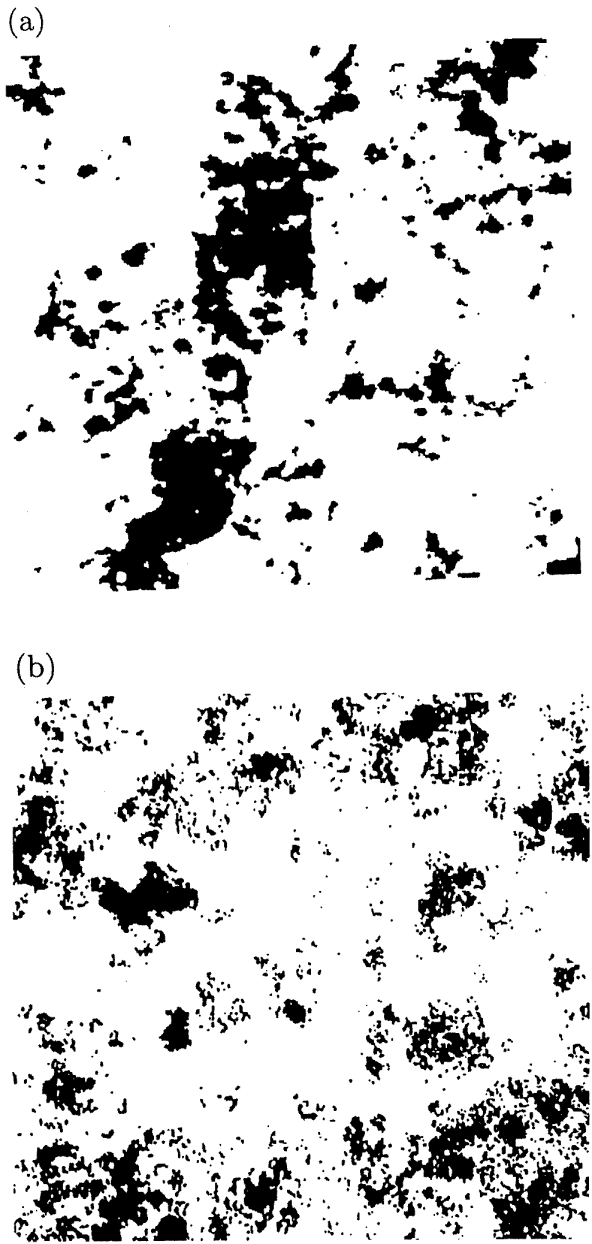

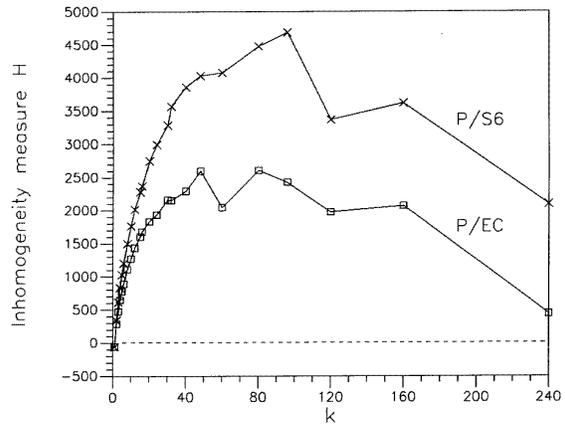

**Figure 9.** A plot of the measures $H$(P/S6) and $H$(P/EC) for a magnification of ×47 000 as functions of different scales $s/\kappa$, where $k = 480/\sqrt{\kappa}$. The broken line represents the level of ideal randomness, $H = 0$.

**Figure 8.** Representative binary images, 480 × 480 pixels, of composites at a magnification of ×47 000. (*a*) P/S6 composite. (*b*) P/EC composite.

It is worth indicating that the inhomogeneity measure $H$ is only one of the possible measures useful in accounting for the influence of spatial distribution on some physical properties. For example, the so-called measure of compactness [15] allows us to evaluate the tendency for the creation of chains of particles in a simple way even in three dimensions. The above measure of compactness is different from the well known cyclomatic ratio allowing for a distinction between compact and ramified clusters [16, 17]. Unfortunately, to the best of our knowledge, there is no accepted definition of the cyclomatic ratio for 3D clusters.

In this paper we have tried to show, the polyethylene/carbon black composites, a relationship between the effective conductivity and the structure of two-phase media by means of some types of Yoshida's geometry. On the basis of Garncarek's quantitative measure of the degree of inhomogeneity, a relevant connection between the same effective property and the 2D representative distribution of carbon black has been observed. The conclusions are as follows.

(a) Both structures P/EC and P/S6 are of the intermediate type in Yoshida's approach. The P/EC structure is considerably close to MG assumptions, while P/S6 shows a mixed type of structure with a tendency towards the BR structure.

(b) The structures of the composites are clearly distinguishable by EMA-Y. Moreover, for P/S6 the tangled structure appears over a larger range of $x_v$ than for P/EC.

(c) The structural differences between the composites are measurable by means of Garncarek's measure $H$ and sensitive changes of the spatial distribution.

(d) Our findings for polyethylene composites with various carbon blacks suggest that there is an interdependence between electrical conductivity and the degree of inhomogeneity of the 2D representative distribution of the carbon black.

## Acknowledgments

We thank Professor M J Grimson from The University of Auckland for his help in improving an earlier version of this paper.## Appendix

The monograph [4] contains constructions of the five measures (inhomogeneity degree, scatter, weak compactness, compactness and strong compactness) of distribution features of a finite set of objects in the limited domain S with the area $s$ (volume $v$), togther with some examples of application of the measures in natural and technical sciences. The method of measurement of changes of interphase surface in a two-phase flow, based on the measures of distribution features of one phase in a second phase, has been proposed. The measure $h$ of inhomogeneity degree has been used to evaluate the perturbation of the unsteady thermal field with a complicated geometrical shape (perpendicular connection of two thick-walled plates). A comparison of the maximum temperature gradient and the measure $h$ leads to the conclusion that $h$ can replace a local physical magnitude, i.e. a temperature gradient in evaluation of the perturbation of thermal fields. To valuate the character of a fluid bed on the basis of the inhomogeneity degree measurement of the arrangement of tracer position in a fluidized bed, an example of application of the measure $H$ has been given. In the second part of the monograph examples of applications of others measures in astronomy and chemical engineering were given. In astronomy the shapes of galaxy agglomerations in Jagiellonian field have been investigated. In chemical engineering the structure evaluation of a two-phase flow has been attempted with respect to the shapes of air bubble agglomerations and with regard to the frequency of their occurrence in comparison to random frequency.